\documentclass{article}
\usepackage{tikz}
\usepackage{bm}
\usepackage{multirow}
\usepackage{booktabs}
\usepackage{amsmath}
\usepackage{amssymb}
\usepackage{graphicx}
\usetikzlibrary{decorations.text}
\newcommand{\arcarrowone}[8]% inner radius, middle radius, outer radius, start angle, end angle, tip protusion angle, options, text
{   \pgfmathsetmacro{\rin}{#1}
    \pgfmathsetmacro{\rmid}{#2}
    \pgfmathsetmacro{\rout}{#3}
    \pgfmathsetmacro{\astart}{#4}
    \pgfmathsetmacro{\aend}{#5}
    \pgfmathsetmacro{\atip}{#6}
    \fill[#7,shift={(40.71mm,25.71mm)}]  (\astart:\rin) arc (\astart:\aend:\rin) -- (\aend+\atip:\rmid) -- (\aend:\rout) arc (\aend:\astart:\rout) -- (\astart+\atip:\rmid) -- cycle;
    \path[decoration={text along path, text={STEP 2: Statistical Analysis}, text align={align=center}, raise=-0.5ex},decorate,shift={(40.71mm,25.71mm)}] (\astart+\atip:\rmid) arc (\astart+\atip:\aend+\atip:\rmid);
}

\newcommand{\arcarrowtwo}[8]% inner radius, middle radius, outer radius, start angle, end angle, tip protusion angle, options, text
{   \pgfmathsetmacro{\rin}{#1}
    \pgfmathsetmacro{\rmid}{#2}
    \pgfmathsetmacro{\rout}{#3}
    \pgfmathsetmacro{\astart}{#4}
    \pgfmathsetmacro{\aend}{#5}
    \pgfmathsetmacro{\atip}{#6}
    \fill[#7,shift={(40mm,24mm)}]  (\astart:\rin) arc (\astart:\aend:\rin) -- (\aend+\atip:\rmid) -- (\aend:\rout) arc (\aend:\astart:\rout) -- (\astart+\atip:\rmid) -- cycle;
    \path[decoration={text along path, text={STEP 3: Update DoE}, text align={align=center}, raise=-0.5ex},decorate,shift={(40mm,24mm)}] (\astart+\atip:\rmid) arc (\astart+\atip:\aend+\atip:\rmid);
}

\newcommand{\arcarrowthree}[8]% inner radius, middle radius, outer radius, start angle, end angle, tip protusion angle, options, text
{   \pgfmathsetmacro{\rin}{#1}
    \pgfmathsetmacro{\rmid}{#2}
    \pgfmathsetmacro{\rout}{#3}
    \pgfmathsetmacro{\astart}{#4}
    \pgfmathsetmacro{\aend}{#5}
    \pgfmathsetmacro{\atip}{#6}
    \fill[#7,shift={(39.29mm,25.71mm)}]  (\astart:\rin) arc (\astart:\aend:\rin) -- (\aend+\atip:\rmid) -- (\aend:\rout) arc (\aend:\astart:\rout) -- (\astart+\atip:\rmid) -- cycle;
    \path[decoration={text along path, text={STEP 1: Perform Experiments}, text align={align=center}, raise=-0.5ex},decorate,shift={(39.29mm,25.71mm)}] (\astart+\atip:\rmid) arc (\astart+\atip:\aend+\atip:\rmid);
}

\DeclareMathOperator*{\argmax}{argmax}

\title{Bayesian Optimization for Intrinsically Noisy Response Surfaces}
\author{Anton van Beek\\ email: anton.vanbeek@ucd.ie}
\begin{document}
\maketitle

While many advanced statistical methods for the design of experiments exist, it is still typical for physical experiments to be performed adaptively based on human intuition. As a consequence, experimental resources are wasted on sub-optimal experimental designs. Conversely, in the simulation-based design community, Bayesian optimization (BO) is often used to adaptively and efficiently identify the global optimum of a response surface. However, adopting these methods directly for the optimization of physical experiments is problematic due to the existence of experimental noise and the typically more stringent constraints on the experimental budget. Consequently, many simplifying assumptions need to be made in the BO framework, and it is currently not fully understood how these assumptions influence the performance of the method and the optimality of the final design. In this paper, we present an experimental study to investigate the influence of the controllable (e.g., number of samples, acquisition function, and covariance function) and noise factors (e.g., problem dimensionality, experimental noise magnitude, and experimental noise form) on the efficiency of the BO framework. The findings in this study include, that the Mat\'{e}r covariance function shows superior performance over all test problems and that the available experimental budget is most consequential when selecting the other settings of the BO scheme. With this study, we enable designers to make more efficient use of their physical experiments and provide insight into the use of BO with intrinsically noisy training data.

\section{INTRODUCTION}
\begin{figure}
\centering
\begin{tikzpicture}
%\draw[step=10mm, black, thin] (0mm,0mm) grid (80mm,50mm);

\node[inner sep=0pt] () at (51mm,32mm)
    {\includegraphics[width=.10\textwidth]{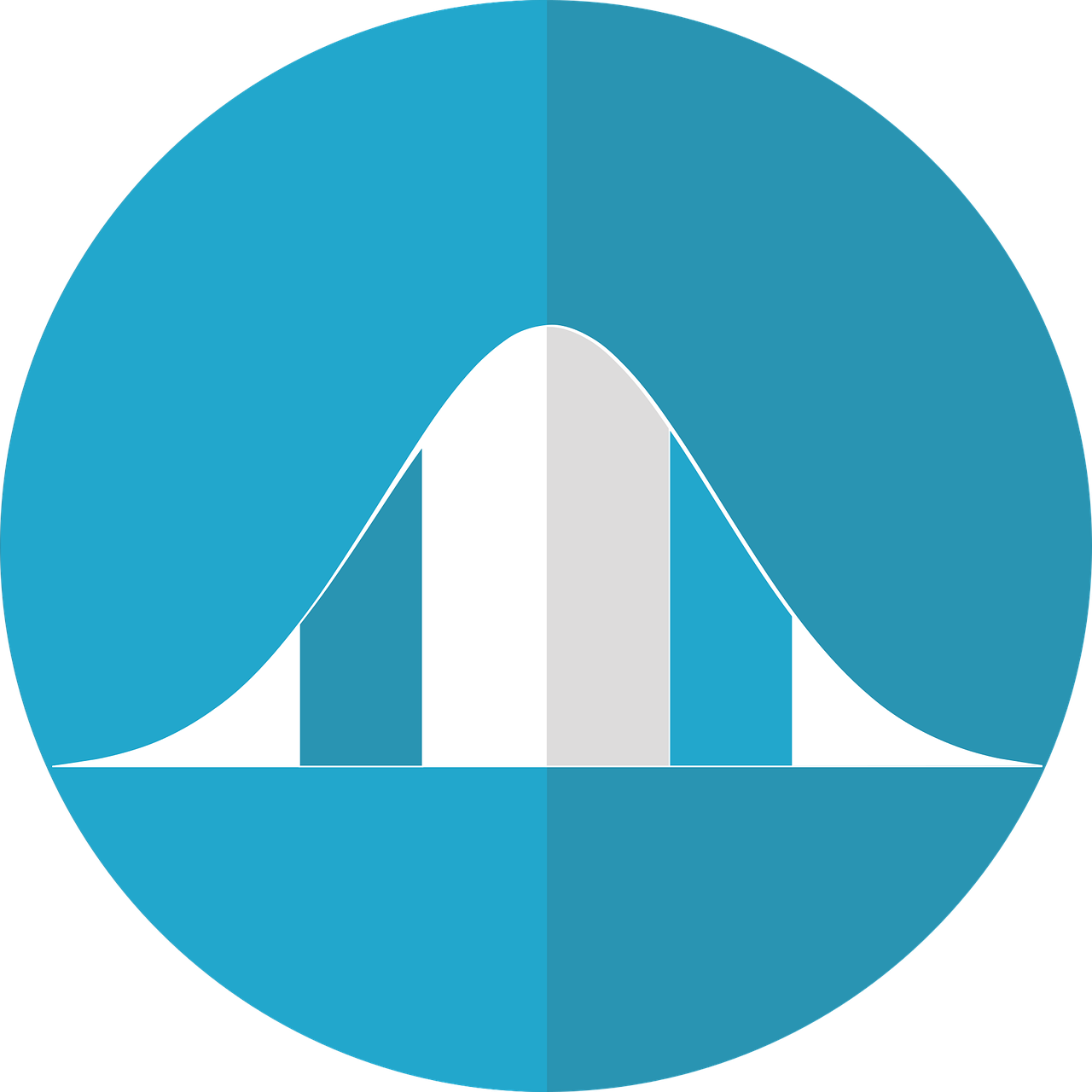}};
\node[inner sep=0pt] () at (28mm,32mm)
    {\includegraphics[width=.14\textwidth]{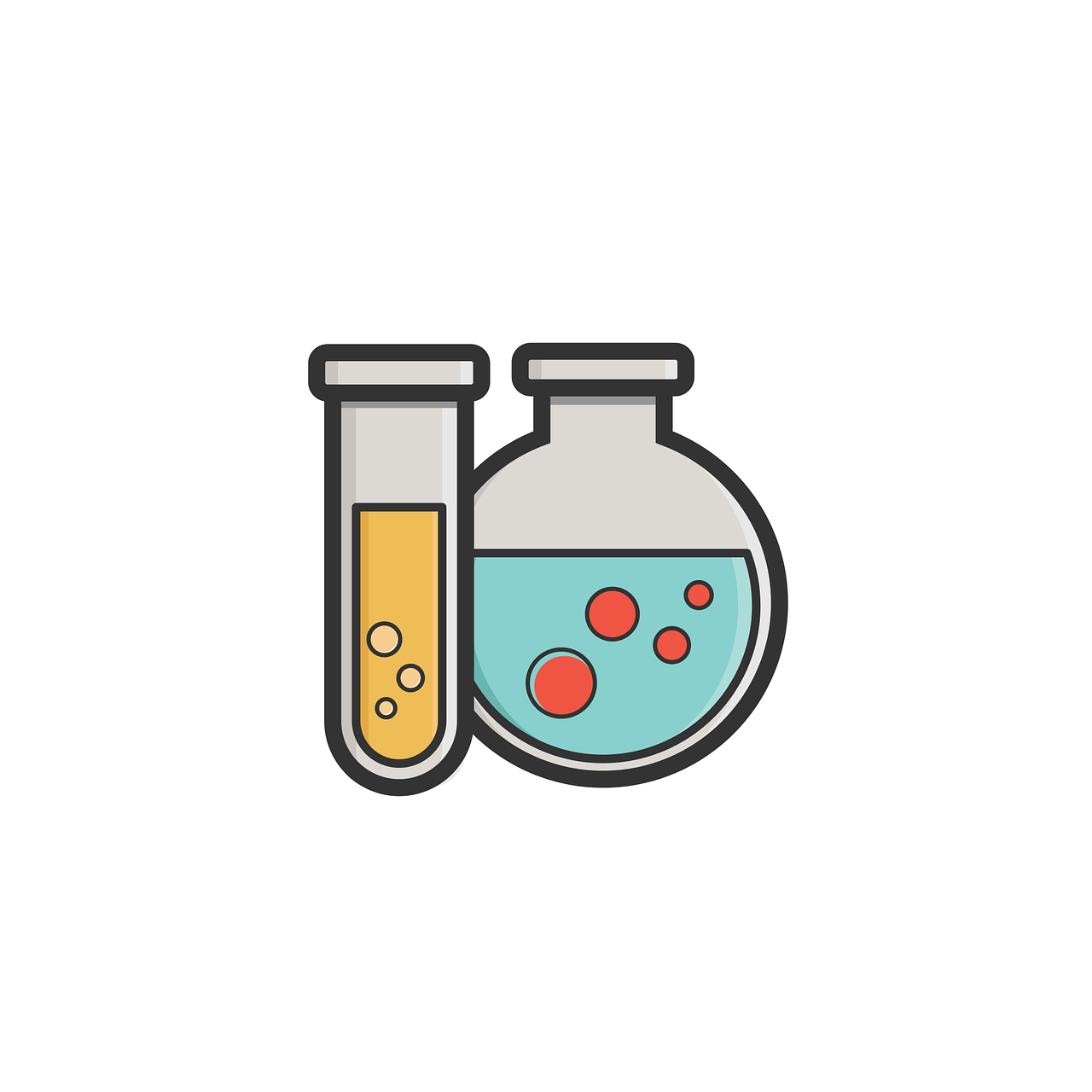}};
\node[inner sep=0pt] () at (41mm,12mm)
    {\includegraphics[width=.10\textwidth]{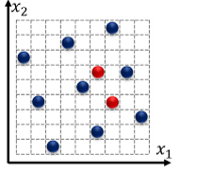}};
\node[inner sep=0pt] () at (9mm,40mm)
    {\includegraphics[width=.10\textwidth]{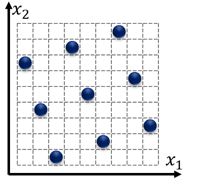}};
\node[inner sep=0pt] () at (70mm,10mm)
    {\includegraphics[width=.14\textwidth]{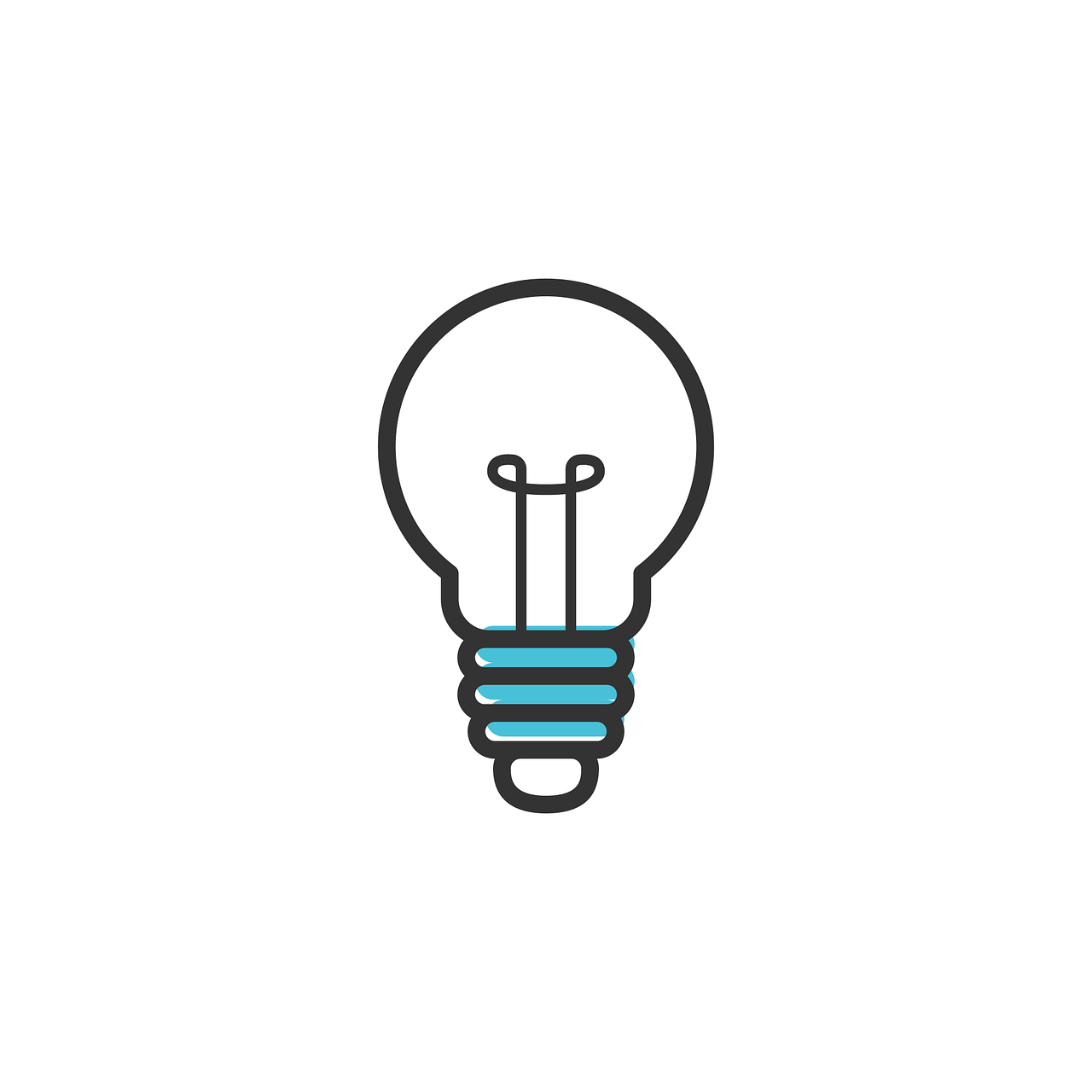}};

\draw[very thick,color = black!50!white,rounded corners=2mm] (13mm,30mm) --++ (-14mm,0mm) --++ (0mm,25mm) --++ (30mm,0mm) --++ (0mm,-6mm) -- cycle;
\fill[very thick,color = black!50!white,rounded corners=2mm] (-1mm,49mm) --++ (0mm,6mm) --++ (30mm,0mm) --++ (0mm,-6mm) -- cycle;
\fill[very thick,color = black!50!white] (-1mm,49mm) --++ (0mm,3mm) --++ (30mm,0mm) --++ (0mm,-3mm) -- cycle;

\draw[line width=1mm,->,-stealth,rounded corners=2mm,color = black!50!white] (6mm,30mm) --++  (0mm,-5mm) --++ (5.5mm,0mm);
\draw[line width=1mm,->,-stealth,rounded corners=2mm,color = black!50!white] (68mm,25.5mm) --++  (5mm,0mm) --++ (0mm,-5.5mm);

\draw[very thick,color = black!50!white,rounded corners=2mm] (51mm,1mm) --++ (0mm,-6mm) --++ (30mm,0mm) --++ (0mm,25mm) --++ (-14mm,0mm) --++ (-4mm,-4mm);
\fill[very thick,color = black!50!white,rounded corners=2mm] (51mm,1mm) --++ (0mm,-6mm) --++ (30mm,0mm) --++ (0mm,6mm) -- cycle;
\fill[very thick,color = black!50!white] (51mm,1mm) --++ (0mm,-3mm) --++ (30mm,0mm) --++ (0mm,3mm) -- cycle;

\draw[very thick] (40.71mm,25.71mm) --++  (90:25mm) arc(90:-30:25mm) -- cycle;
\draw[very thick] (40mm,24.5mm) --++  (-30:25mm) arc(-30:-150:25mm) -- cycle;
\draw[very thick] (39.29mm,25.71mm) --++  (-150:25mm) arc(-150:-270:25mm) -- cycle;

\arcarrowone{2.25}{2.5}{2.75}{95}{-25}{-5}{red!50!white,draw=red!50!white,thin}{};
\arcarrowtwo{2.25}{2.5}{2.75}{-25}{-145}{-5}{green!50!white,draw=green!50!white,thin}{};
\arcarrowthree{2.25}{2.5}{2.75}{-145}{-265}{-5}{blue!50!white,draw=blue!50!white,thin}{};

\path (3mm,52mm) node[anchor=west]{Initial DoE};
\path (54mm,-2mm) node[anchor=west]{Final Design};

\end{tikzpicture}
\caption{BO for the adaptive design of physical experiments, a generalized framework. In the iterative process between the acquisition of data and statistical analysis.}
\label{introfig}
\end{figure}
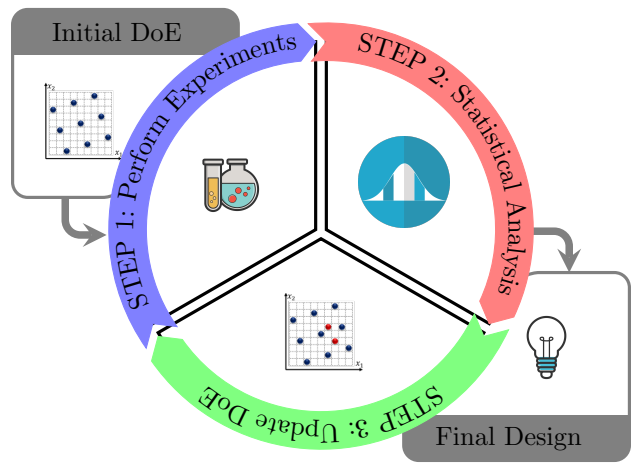

While Bayesian optimization (BO) is a well-established method for data efficient optimization of deterministic and time intensive response surfaces \cite{Forrester2009}, its application for the optimization of stochastic response surfaces is still an elusive objective. Stochastic response surfaces are functions that manifest intrinsic uncertainty so that when they are evaluated for the same inputs a variation in the output is observed. These types of response surfaces are encountered when doing physical experiments (e.g., graphene exfoliation \cite{chaney2024,hui2024}) and some forms of simulation experiments (e.g., molecular dynamics \cite{Beek2021}, and agent-based models). One primary advantage of BO is that it provides a systematic approach to sequentially identify the next most appropriate input conditions to evaluate, thus saving experimental resources. This approach is shown in Figure~\ref{introfig}, where it can observe that the process starts with a small number of uniformly distributed samples. Subsequently, we have an iterative process that involves response surface approximation, identification of a new input condition, and then experimentation. Once a stopping condition has been reached this process is terminated. Moreover, the sampling decisions are made by maximizing an acquisition function that balances the mean of a posterior predictive distribution (i.e., a Bayesian-based response surface approximation conditioned on the available training data set) with the interpolation uncertainty (i.e., exploitation versus exploration).

The typical approach for the exploration of unknown response surfaces is the use of an experimental design that involves a set of a uniformly distributed set of input conditions \cite{Santner2003, Box1964}. Examples of such experimental designs are factorial designs \cite{Mukerjee2006}, Latin hypercube samples \cite{Jin2005}, and Sobol sequences \cite{Sobol1976}. These approaches have been applied for the design of experiments (DoEs) in the chemical sciences \cite{Igder2012}, civil engineering \cite{Durakovic2017}, and psychology \cite{Lindquist1953}. While these approaches have been shown to be more data efficient than studying one factor at a time \cite{Fisher1992}, they do not enable the designer to leverage the information of previous experiments to inform subsequent experiments. Consequently, BO provides a compelling, and potentially more data-efficient, alternative to one-shot DoEs. 

The challenge when using BO to optimize stochastic response surfaces lies in the need to obtain a posterior predictive distribution, of the unknown response surface, that can quantify the intrinsic data uncertainty from the interpolation uncertainty. However, establishing such a posterior predictive distribution is a data-intensive task, and thus simplifying assumptions often need to be made. Examples of such methods include Practical Kriging \cite{Binois2018}, Stochastic Kriging \cite{Ankenman2008}, and Gaussian process (GP)-based quantile regression \cite{Plumlee2014}. However, stochastic Kriging and GP-based quantile regression require many replicates to learn the form of the experimental uncertainty, whereas practical Kriging involves placing an additional GP on the experimental variance that is data-intensive to learn.

In this paper, we present an empirical investigation into the effect that decisions in the construction of the posterior predictive distribution (e.g., choice of covariance function, acquisition function, the use of replicates, and initial batch size) and the properties of the response surface (e.g., noise magnitude, problem dimension, and noise form) have on the efficiency of the BO process. Through this effort we can make the following two knowledge claims about BO in the context of stochastic response surfaces:
\begin{enumerate}
    \item The obtained insight enables engineers and scientists to use prior knowledge of the properties of the response surface to inform the construction of the posterior predictive distribution in the BO framework.
    \item It highlights under what conditions it is appropriate to use the BO framework for the optimization of stochastic functions. 
\end{enumerate}
Finally, we will use the presented study to explore how the BO framework can be improved to be more appropriate for the optimization of stochastic functions. Through this effort, we hope to empower designers of physical experiments (e.g., engineers and scientists) to benefit from advanced statistical tools and get a better understanding of the knowledge embedded in the processes that they study.

\section{STUDY BACKGROUND}
In this section, we will introduce the methods used to approximate stochastic response surfaces, available acquisition functions, and the configuration of the experimental study presented in this work. 

\subsection{Gaussian Process Modeling}
While multiple forms of response surfaces have been used for adaptive optimization of costly to evaluate objective functions (e.g., neural networks \cite{Springenberg2016}), in this work we will only be using GPs. The reason is their ease of implementation and generalization to a plethora of different problems. Assume that we have a set of $n$ noisy observations $\textbf{Y}=\left\{y_1,\ldots,y_n\right\}^T$ for a set of $d$-dimensional input conditions $\textbf{X}=\left\{\textbf{x}_1,\ldots,\textbf{x}_n\right\}^T$. Consequently, we would like to establish an emulator on the function $f:\mathbb{R}^d\rightarrow \mathbb{R}$. Under the assumption that the observations are jointly normally distributed, we can place a GP prior on the unknown response surface $f$ and characterize it through a mean function and a covariance function $k:\mathbb{R}^d\times \mathbb{R}^d\rightarrow \mathbb{R}$ \cite{Williams2006}. Consequently, an observation model can be established as $y(\textbf{x}_i)=f(\textbf{x}_i)+\varepsilon_i$ where the noise is normally distributed as $\varepsilon_i\sim\mathcal{N}(0,\mathcal{S}(\textbf{x}_i))$. Alternatively, we can write $Y \sim \mathcal{N}_n\left(\textbf{M}\bm{\beta}^T, \textbf{K}_n + \boldsymbol{\Sigma}_n\right)$, where $\textbf{K}_n$ is a $n\times n$ matrix with the $ij$ coordinates given as $k(\textbf{x}_i,\textbf{x}_j)$, $\bm{\Sigma}_n=\text{diag}\left(\mathcal{S}(\textbf{x}_1),\ldots,\mathcal{S}(\textbf{x}_n) \right)$ is a diagonal matrix that accounts for the experimental uncertainty, $\textbf{M}$ is a $n\times p$ matrix where the $i^{th}$ row is a vector of $p$ basis vectors given as $\textbf{m}(\textbf{x})=\left\{ m_1(\textbf{x}),\ldots,m_p(\textbf{x}) \right\}^T$.%, and $\textbf{B}$ is an $n\times p$ matrix where the $i^{th}$ row is a vector of $p$ weights given as $\bm{\beta} = \left\{ \beta_1,\ldots,\beta_p \right\}$. 

Concerning the covariance structure of the GP, under the assumption of \textit{homoscedastic} noise (i.e., $\mathcal{S}(\textbf{x})=\tau$), the experimental uncertainty can be defined by a single variable such that $\bm{\Sigma}_n=\tau\textbf{I}_n$ where $\textbf{I}_n$ is an $n$-dimensional identity matrix. While GPs have been extended to be applicable to data with \textit{heterscedastic} noise (e.g., Practical Kriging \cite{Binois2018}, Stochastic Kriging \cite{Ankenman2008}, and GPs based quantile regression \cite{Plumlee2014}), they are impractical for the optimization of physical experiments with small computational budgets (e.g., less than $10 d$ samples \cite{Beek2021}). In addition, the correlation between observations is accounted for through the covariance function $k(\cdot,\cdot)$, which is often selected as the squared exponential that is defined as
\begin{align}
k(\textbf{x},\textbf{x}^{\prime}) & =\sigma^2\exp\left( \sum_{i=1}^d -10^{\omega_i}(x_i - x_j^{\prime})^2\right),\nonumber\\
& = \sigma^2r(\textbf{x},\textbf{x}^{\prime}),
\end{align}
where $\sigma^2$ is the prior variance and $\bm{\omega}=\left\{\omega_1,\ldots,\omega_d\right\}^T$ is the roughness of the response surface.

Under these conditions, we can approximate the model parameters by maximizing their log-likelihood profile as
\begin{equation}\label{llh}
\bm{\hat{\omega}}.\hat{\tau} = \argmax_{\bm{\omega},\tau\in\Omega\times T}  - n \log\left( \hat{\sigma}^2\right) - \log\left(\left|\textbf{V}\right|\right),
\end{equation}
where $\textbf{V} = \textbf{R}_n + \tau\textbf{I}_n$, the $i,j^{th}$ element of $\textbf{R}_n$ is $r(\textbf{x}_i,\textbf{x}_j)$, and the constant terms have been dropped. In addition, the search space has been defined as $\Omega\in \left[-10, 10 \right]^d$ and $T\in \left[0,1 \right]$ (this is reasonable when normalizing the training data). Note that the consideration of experimental noise adds only a single additional hyperparameter to infer. Moreover, taking the derivative of the likelihood we can solve for $\hat{\bm{\beta}}$ and $\hat{\sigma^2}$ as
\begin{align}
\hat{\bm{\beta}} &= \left(\textbf{M}^T \textbf{V}^{-1}\textbf{M}\right)\textbf{M}^T\textbf{V}^{-1}\textbf{Y},\\
\hat{\sigma^2} &= \frac{1}{n}\left(\textbf{Y} - \textbf{M}\hat{\bm{\beta}}\right)^T \textbf{V}^{-1}\left(\textbf{Y} - \textbf{M}\hat{\bm{\beta}}\right).
\end{align}
As an alternative to maximum likelihood estimation, the designer can use the maximum a posteriori probability, and cross-validation to get a points estimate of the hyperparameters or use a Bayesian approach to account for the uncertainty in the hyperparameters \cite{Williams2006}. While the Bayesian approach has shown to be superior in performance \cite{De2021}, we rely on the maximum likelihood approximations of the parameters for its numerical stability and computational efficiency \cite{ Martin2005}.

Having a point estimate of the hyperparameters enables a designer to condition the prior distribution of the unknown response surface on the observed data $\textbf{D}=\left\{\textbf{X},\textbf{Y}\right\}$. Specifically, the posterior approximation $Y(\textbf{x})|\textbf{D}$ for input inputs conditions $\textbf{x}$ are fully defined through its mean and variance as
\begin{align}
\mu(\textbf{x})&=\textbf{m}(\textbf{x})\hat{\bm{\beta}} + \textbf{k}(\textbf{x})^T\Lambda^{-1}(\textbf{Y}-\textbf{M}\hat{\bm{\beta}}),\\
s^2(\textbf{x})&=k(\textbf{x},\textbf{x}) - \textbf{k}^T(\textbf{x})\Lambda^{-1}\textbf{k}(\textbf{x})\nonumber\\&\qquad  +\textbf{W}^T\left(\textbf{M}^T \Lambda^{-1}\textbf{M}\right)^{-1}\textbf{W}+\hat{\sigma}^2\hat{\tau} ,
\end{align}
respectively. Moreover, $\textbf{W}=\textbf{m}(\textbf{x})-\textbf{M}^T\Lambda^{-1}\textbf{k}(\textbf{x})$, $\Lambda=\hat{\sigma}^2\textbf{V}$, and $\textbf{k}(\textbf{x})$ is an $n\times 1$-dimensional vector whose $i^{th}$ element is given as $k(\textbf{X}_i,\textbf{x})$. This posterior approximation has the advantage that it provides a quantification of the prediction uncertainty; however, it is not readily salient how it can be used to identify new input conditions to test.

\subsection{Acquisition Functions}
A wide variety of acquisition functions have been proposed in the literature; however, none of them have proven to universally outperform the others \cite{Forrester2009}. These acquisition functions are used to identify what input conditions $\textbf{x}^{new}_t$ should be tested next at step $t$ of the optimization process according to
\begin{equation}
\textbf{x}^{new}_t = \argmax_{\textbf{x}\in\chi} \alpha(\textbf{x}|\textbf{D}),
\label{general}
\end{equation}
where $\chi$ is space of admissible input conditions, and $\alpha(\cdot)$ is the selected acquisition function. In this subsection, we will introduce a set of five alternative acquisition functions that are investigated in this study.

\begin{figure*}[t]
\includegraphics[width=\textwidth,trim={2.95cm 0 1.85cm 0},clip]{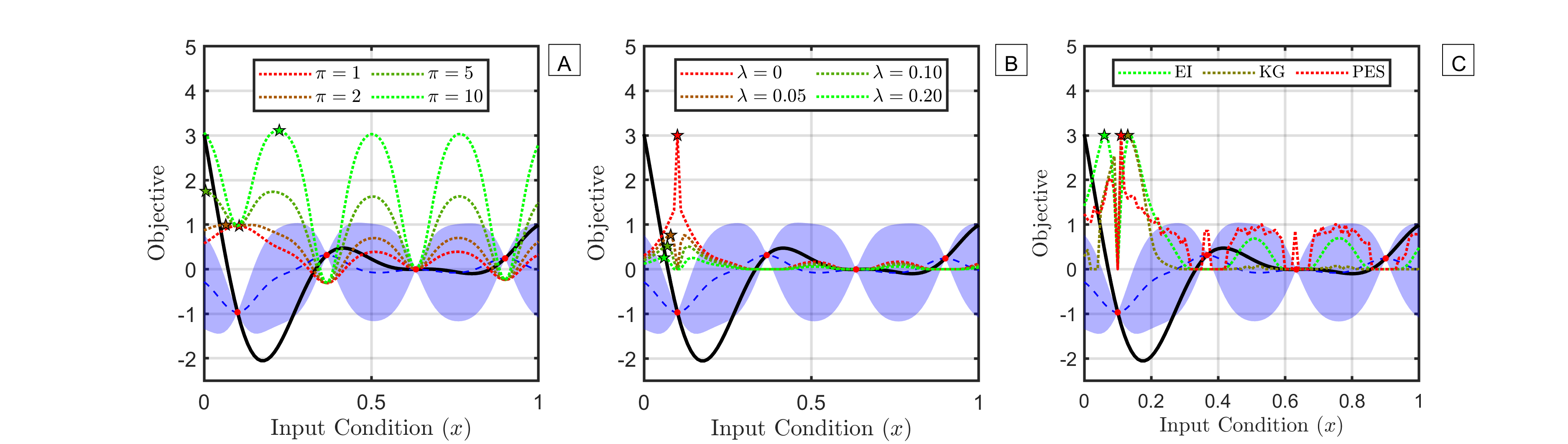}
\caption{Visualization of acquisition functions for an arbitrary objective function (black line) that has been approximated with a GP (blue dashed line shaded region), trained on four samples (red dots). A) scaled UC for four values of $\pi = \left\{1, 2, 5, 10 \right\}$, B) normalized PI for four different values of $\lambda = \left\{0, 0.05, 0.10, 0.20 \right\}$, and C is the normalized EI, KG, and PES.) The stars indicate the optimal and new sampling locations for each acquisition function.}
\label{AQplots}
\end{figure*}

The first acquisition function that we will introduce is the statistical lower bound \cite{Shahriari2015}. This is considered a nonrigorous branch-and-bound algorithm that involves minimizing $\mu(\textbf{x}) - \pi s(\textbf{x}))$ where for any value of $\pi > 0$ we have the desired property of balancing exploration with exploitation \cite{Jones2001}. To be consistent with the optimization formulate given in Equation~\ref{general}, we reformulate this expression as
\begin{equation}
    \alpha_{UC} (\textbf{x}) =\pi s(\textbf{x}) + \mu(\textbf{x}),
\end{equation}
and refer to this as the upper confidence (UC) acquisition function throughout the remainder of this paper. While the UC objective is intuitively appealing, it is known to exclude regions of the space of admissible input conditions, and thus does not guarantee the necessary sample density requirement to ensure convergence to a global optimum \cite{Torn1989}. In panel A of Figure~\ref{AQplots} we show the UC for four different values of $\pi$, from which it can be observed that more emphasis is placed on exploration for larger values of $\pi$.  

The second acquisition function that we will investigate is the \textit{probability of improvement} (PI)\cite{Jones2001}. Improvement in this method is defined as $I(\textbf{x})=\max(\left\{y_t^* - Y(\textbf{x}), 0 \right\})$, where $y_t^*$ is the best observed sample at the $t^{th}$ iteration (i.e., $y_t^* = \min(\textbf{Y})$). Given that $y(\textbf{x})$ is a random variable, it becomes possible to calculate the PI through integration of the improvement from negative infinity up to the current best observed sample $y_t^*$. However, using this directly as an acquisition function has shown to place too much emphasis on exploitation, making it susceptible to waste experimental resources on improving the response surface accuracy around local optima. Consequently, the typical implementation of the PI is defined as
\begin{equation}
\alpha_{PI} (\textbf{x}) = \Phi\left(\frac{y_t^{(tar)} - \mu(\textbf{x})}{s(\textbf{x})}\right),
\end{equation}
where $y_t^{(tar)} = y_t^* - \lambda\left(\max(\textbf{Y}) - y_t^*\right)$ is the target response surface value, and $\Phi(\cdot)$ is the standard normal cumulative density function. In this case, designers can place more emphasis on exploration by selecting larger values for $\lambda$. For visualization of the PI, in panel B of Figure~\ref{AQplots} we have plotted the PI for a test function with respect to five different values of $\lambda=\left\{0, 0.05, 0.10, 0.2, 0.5 \right\}$ (scaled to be in a range of 0 to 3). Observe how the new sampling locations change more towards exploring unobserved regions for higher values of $\lambda$.

An alternative to the PI function is the \textit{expected improvement} (EI) function. As the name suggests, it involves taking the expectation of the improvement as $\mathbb{E}\left(\max(\left\{y_t^* - Y(\textbf{x}), 0 \right\})\right)$. However, in the case of the stochastic response surface, the existence of experimental uncertainty places too much emphasis on exploitation. Consequently, \cite{Huang2006} proposed a modified version of the EI function as
\begin{align}
\label{EI}
\alpha_{EI}(\textbf{x}) & = \mathbb{E}\left(\max(\left\{y_t^* - Y(\textbf{x}), 0 \right\})\right)\left(1 - \frac{\hat{\sigma}^2\hat{\tau}}{s(\textbf{x})} \right),\nonumber \\
& =  s(\textbf{x})\left( u \Phi(u) + \phi(u) \right)\left(1 - \frac{\hat{\sigma}^2\hat{\tau}}{s(\textbf{x})} \right),
\end{align}
where $\phi(\cdot)$ is the standard normal probability density function and $u=\frac{y_t^* - \mu(\textbf{x})}{s(\textbf{x})}$. Note that $s(\textbf{x})$ equals $\hat{\sigma}^2\hat{\tau}$ for any $\textbf{x}\in\textbf{X}$. For visualization purposes, we have plotted the modified EI function Equation~\ref{EI} in panel C of Figure~\ref{AQplots}. Observe how the conventional EI is nonzero for observed samples and as such new samples are more likely to be allocated close to the current observed best sample $y_t^*$.

The next acquisition function that we will test is the \textit{knowledge gradient} (KG) that was first introduced in \cite{Frazier2008}. The KG is an acquisition function that involves a one-step look-ahead policy that aims to maximize the difference between the optimal response $\min{\left(Y(\textbf{x})|\textbf{D}_t\right)}$ at step $t$ and the optimal response after observing $\textbf{x}_t^{(new)}$ \cite{Frazier2012}. Specifically, the KG is defined as
\begin{align}
\alpha_{KG}(\textbf{x})&=\mathbb{E}\left(Y(\textbf{x}_t^*)|\textbf{D}_t \right. \nonumber \\ &\quad- \left.Y(\textbf{x}_{t+1}^*)|\textbf{D}_{t+1},\textbf{x}_t^{new}=\textbf{x}\right), 
\end{align}
where the subscripts on $\textbf{D}_t$ have been used to indicate that new observations have been added to the initial training data set (i.e., $\textbf{D}_t=\bigcup_{i=1}^t\left\{x_i^{new}, y_i^{new} \right\}\cup\textbf{D}$). However, it should be noted that for $\textbf{D}_{t+1}$ we have not yet observed the response $y_{t+1}^{new}$, and thus we need to take the expectation with respect to its predicted value \cite{Tao2021}. In panel C of Figure~\ref{AQplots} we have plotted the KG for a test function, and show its slight difference from the other acquisition functions.

The final accustoming function that we will introduce and investigate is the \textit{predictive entropy search} (PES) that was first introduced in \cite{Hernandez2014}. Instead of trying to maximize the improvement of the KG, the PES uses information theory to maximize learning about the spatial location of the globally optimal response. Specifically, the acquisition function is defined as
\begin{align}
\label{PES}
\alpha_{KG}(\textbf{x})&=H\left(Y(\textbf{x})|\textbf{D}_t \right) \\
&\quad- \mathbb{E}_{p(\textbf{x}^*|\textbf{D}_t)}\left(H(Y(\textbf{x}_{t+1}^*)|\textbf{D}_{t+1},\textbf{x}_t^{new}=\textbf{x}) \right) \nonumber,
\end{align}
where $H(p(\textbf{x})) = -\int p(\textbf{x})\log p(\textbf{x})d\textbf{x}$, and $\mathbb{E}_{p(\textbf{x}^*|\textbf{D}_t)}$ is the probability that the global optimum is found at input condition $\textbf{x}$ after observing observations $\textbf{D}_t$. The first term on the left-hand side of Equation~\ref{PES} has a closed-form expression, whereas the right-hand side must be approximated through, for example, expectation propagation \cite{Minka2001}. Similar to previous acquisition functions, we have plotted the PES in panel C of Figure~\ref{AQplots}.

\subsection{Experimental Setup}
In this subsection, we will discuss the setup of the experimental setup used to investigate the importance of alternative optimization conditions. For this purpose, we have made a distinction between factors that are controllable and a set of uncontrollable noise factors as shown in Table~\ref{xperimental settings}. The purpose of this distinction is that prior knowledge of the noise factors could inform what levels to choose for the controllable factors.

\begin{table*}[t]

  \centering
  \caption{Selected factors and their associated levels under which the experiments have been performed.}
  \label{xperimental settings}
  \begin{footnotesize}
  \begin{tabular}{p{6mm}   p{13mm}  p{11mm}  p{12mm}  p{12mm} p{.1mm} p{8mm} p{14mm} p{10mm} }\\
  \toprule
  & \multicolumn{4}{c}{Controllable Factors} && \multicolumn{3}{c}{Noise Factors}\\
   \cline{2-5}  \cline{7-9}

  Level & Initial replicates & Initial samples & Acquisition function & Covariance function && Problem & Noise magnitude & Noise form \\

  \midrule
  
  1 & 1 & $2d$ & UCB& Gaussian    && $f_1(\cdot)$ & $0.01\Delta_f$ & Constant \\
  2 & 2 & $5d$ & PI & Power       && $f_2(\cdot)$ & $0.05\Delta_f$ & Bad \\
  3 & 3 & $10d$& EI & Mat\'ern    && $f_3(\cdot)$ & $0.20\Delta_f$ & Good \\
  4 &   &      & KG &             &&                  &                   &  \\
  5 &   &      & PES&             &&                  &                   &  \\
  \bottomrule
  \end{tabular}
\end{footnotesize}
\end{table*}

We considered four controllable factors, the initial number of replicates, the initial number of samples, the acquisition function, and the selected covariance function. The initial number of samples is the unique sampling location in the initial DoE. However, for small sample sizes, it might be found that no accurate approximation of the experimental variance $\sigma^2\tau$ can be obtained. Consequently, we also considered the scenario of having an initial number of replicates (i.e., for the same unique initial inputs the response surface function is evaluated 1, 2, or 3 times). In addition, we considered the use of all five previously introduced acquisition functions where for the UC and PI, we set $\pi=5$ and $\lambda=0.1$, respectively. Finally, we considered three different types of covariance functions, the previously introduced Gaussian covariance, the power exponential covariance function that are defined as
\begin{equation}
k(\textbf{x},\textbf{x}^{\prime})  =\sigma^2\exp\left( \sum_{i=1}^d -10^{\omega_i}(x_i - x_j^{\prime})^p\right),
\end{equation}
and the Mat\'{e}rn covariance function that is defined as
\begin{equation}
k(\textbf{x},\textbf{x}^{\prime})  =\sigma^2\frac{2^{1-v}}{\Gamma(v)}\left(\sqrt{2v}\frac{d}{\rho}\right)^v K_v\left(\sqrt{2v}\frac{d}{\rho} \right),
\end{equation}
where $p,\rho,v$ are additional hyperparameters, $K_v(\cdot)$ is the modified Bessel function of the second kind, and $\Gamma(\cdot)$ is the gamma function. The consideration of additional power exponential and Mat\'{e}rn covariance is of interest as they allow more freedom in the form of the covariance but involves additional parameters. Consequently, considering their choice in this study will help determine under what conditions using more complex covariance functions will be beneficial.

\begin{figure*}[t]
\centering
\includegraphics[width=\textwidth]{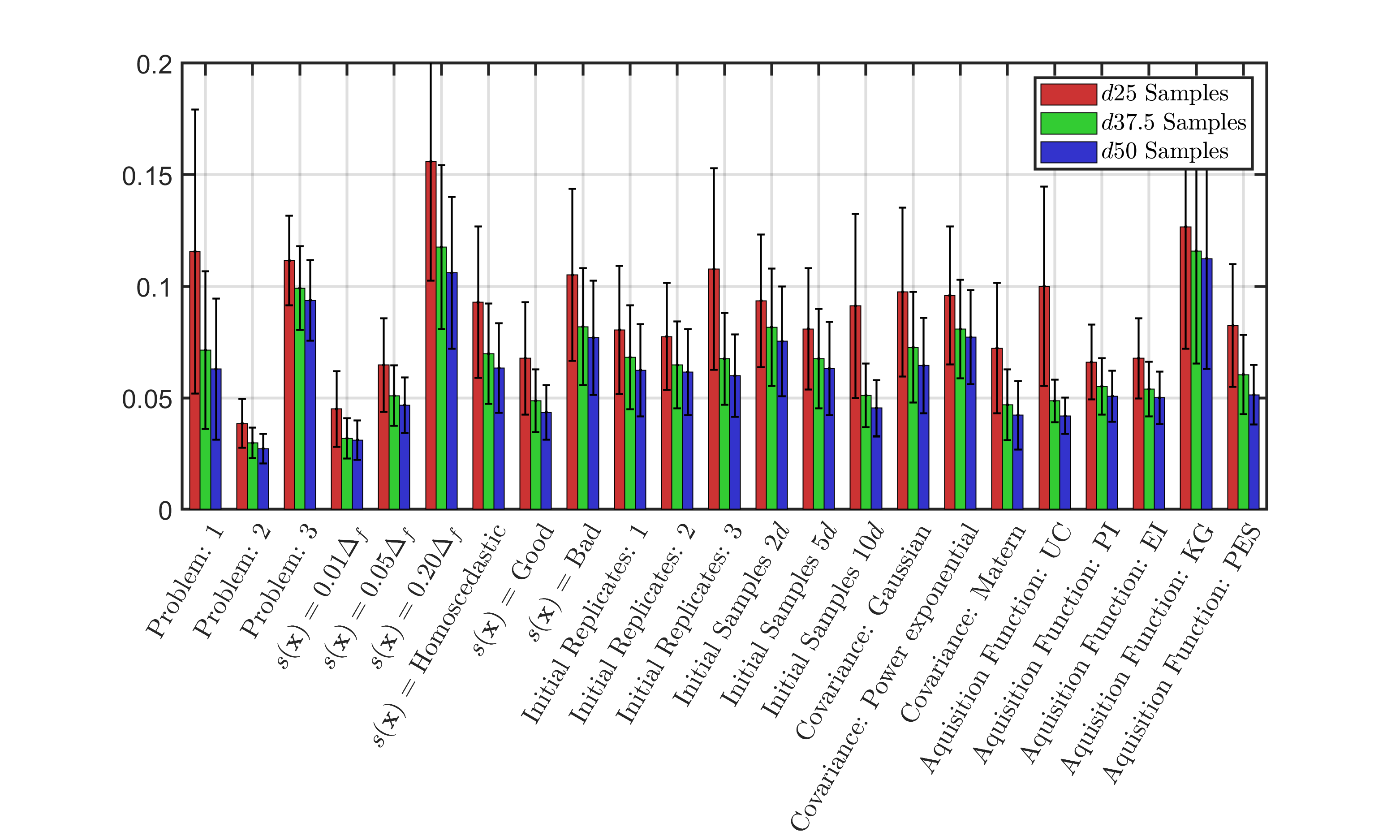}
\caption{The main effects of the controllable and noise factors as measured for the GAP obtained after observing $25d$ (red bars), $37.5d$ (green bars), and $50d$ samples (blue bars).}
\label{maineffect}
\end{figure*}

Concerning the noise factors, we considered the magnitude of the noise and form of the noise. Specifically, we considered three noise scenarios where the maximum standard deviation of the noise is either $\left\{2, 5, 20 \right\}\%$ of the range of the function $\Delta_f$. In addition, we considered three types of experimental uncertainty, one where the variance of the response surface is homoscedastic, one where the noise form is disadvantaged as the variance is maximized at the minimum of the response surface (referred to as bad), and one advantage where the variance is minimized at the global objective (referred to as good). This has been achieved through the formulation $\varepsilon(\textbf{x})\sim \mathcal{N}\left(0, \mathbb{E}(f(\textbf{x}) + b)a \right)$, where parameter $a$ and $b$ have been selected to ensure that the standard deviation ranges from 0.25 to 1.6 times the selected noise magnitude \cite{Beek2021}. Finally, we considered the following set of noisy objective functions

\begin{align}
f_1(x) &= (3x-2)^2sin(12x-4)+\varepsilon, \\
f_2(\textbf{x}) &=\frac{1}{51.95}\left(15x_2 - \frac{5.1(15x_1 -5)^2}{4\pi^2} + \frac{5(15x_1 - 5)}{\pi} -6\right)^2\nonumber \\& \quad\quad +\frac{1}{51.95}\left(\left( 10 - \frac{10}{8\pi}  \right)cos(15x_1-5) - 44.81\right)+\varepsilon, \\
f_3(\textbf{x}) &=4x^2_1 - 2.1x_1^4+\frac{x_1^6}{3}+x_1 x_2 - 4x_2^2 + 4x_2^4 + \varepsilon, 
\end{align}

where $\varepsilon\sim\mathcal{N}(0,\mathcal{S}(\textbf{x}))$. In addition, the spaces of admissible input conditions are $x\in(0,1)$, $\textbf{x}\in(0,1)^2$, and $\textbf{x}\in(-2,2)\times(-1,1)$ for $f_1(\cdot)$, $f_2(\cdot)$, and $f_3(\cdot)$, respectively. Note that we have used only low-dimensional problems as optimizing high-dimensional objective functions will be practically infeasible with small experimental budgets. 

From the above set of factors, we are able to identify a total of $3^6\times 5 = 3645$ unique experiments. Finally, to account for the potential variability associated with random initial conditions, we have repeated all experiments five times for a total of 18225 experiments.

\section{RESULTS AND INTERPRETATION}
In this section, we will study the data obtained from the experiments delineated in the previous section. Specifically, we aim to find what the main effects are of each individual factor and then try to identify the interaction effects between the controllable and noise factors to help guide modeling decisions.

\subsection{Main Factor Effects}
The main effects in this study have been measured by taking the average difference between the identified global optimum and the true global optimum for each factor. This metric is referred to as the GAP \cite{Picheny2013}. A bar chart of these results has been plotted in Figure~\ref{maineffect} where the red bars indicate the GAP after observing the first $25d$ samples, the green bars indicate the GAP after observing the first $37.5d$ samples, and the blue bars indicate the gap after observing all $50d$ samples.

What can be observed from Figure~\ref{maineffect} is that the initial number of replicates has only a minor effect on the optimality of the result obtained from the optimization process. Except, as fewer total samples have been observed, we find that the GAP becomes larger. This is intuitively sensible because when only a small number of samples have been observed then replication of experiments results in less coverage of the admissible design space. Conversely, when looking at the number of initial samples, we find that a large set of initial experiments has a positive effect on the GAP of the final result. What this implies, is that the tendency of the acquisition functions studied in this paper is that they emphasize the exploitation over exploration. Concerning the choice of covariance function, we find that the power exponential performs worst, while the mat\'{e}r covariance performs significantly better. This could suggest that the freedom offered by the mat\'{e}r covariance is beneficial to the optimization process. However, it is interesting to note that this is not the case for the power exponential. The reason for this might be that the Mat\'{e}r covariance provides significantly more modeling freedom compared to the power exponential. Finally, concerning the acquisition function, it is interesting to observe that the UC acquisition function works best for large data scenarios (e.g., $50d$) whereas, in low data scenarios, (e.g., $25d$) the PI and EI appear to have similar performance. The reason for this might be that we used a relatively large value for $\pi = 5$ that significantly emphasizes exploration. Consequently, the main effects of the controllable factors suggest that conventional acquisition place too much emphasis on exploitation and too little on exploration. 

Concerning the main effects of the noise factors, we observed intuitively sensible results. Specifically, we find that the relatively linear problem 2 has a small average GAP, whereas the opposite is observed for problem 2 which has six local minima. In addition, functions with higher magnitudes of noise perform worst in the observed final GAP. Finally, functions with heteroscedastic noise that are minimum at the global optimum perform best. While these insights provide little novelty in terms of insight, they do provide validation of the performed study.

\begin{figure}
\centering
\includegraphics[width=0.8\textwidth]{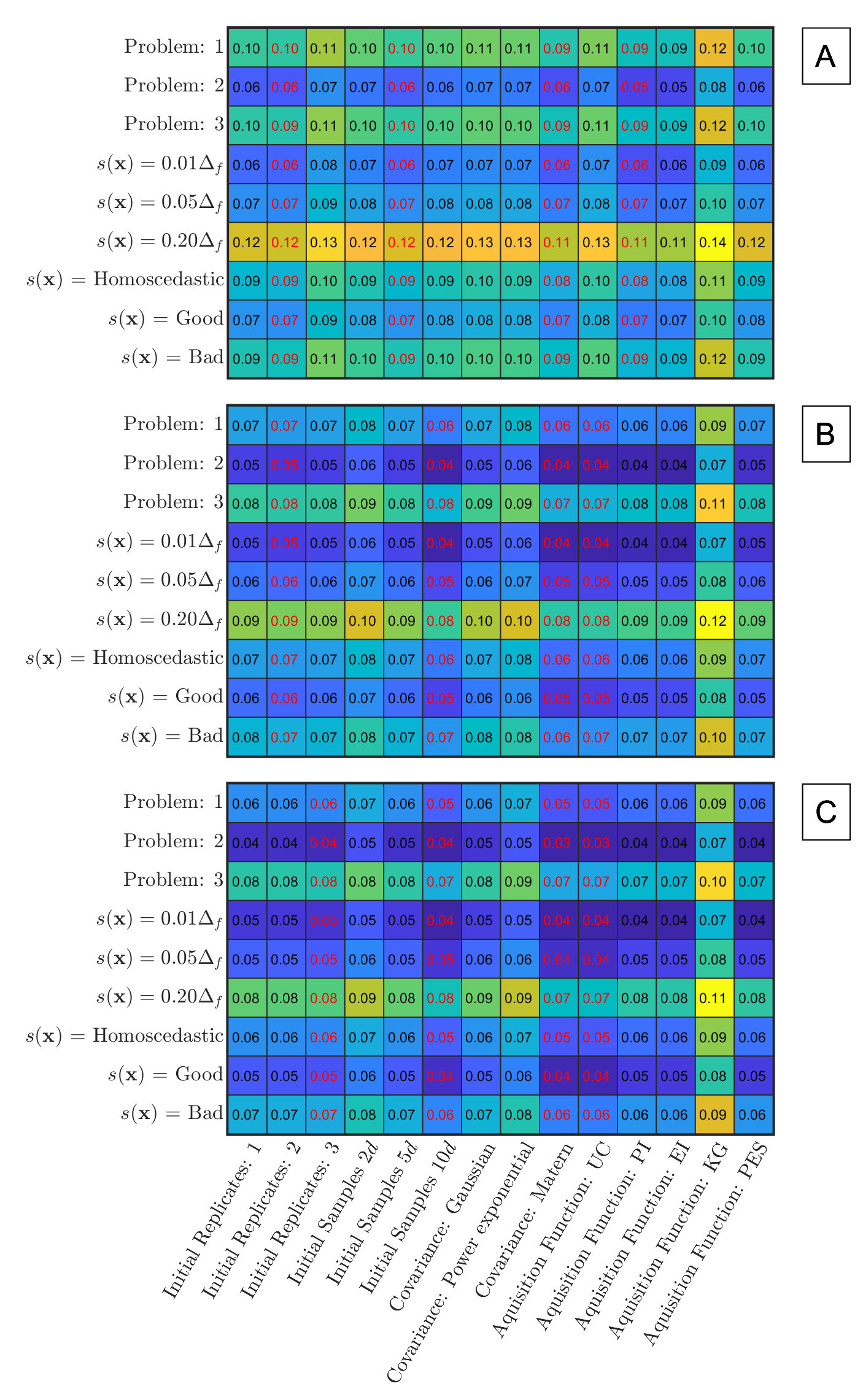}
\caption{The GAP averaged (numbers in each block) over the simulations for all combinations of controllable (horizontal axis) and noise factors (vertical axis), where the red numbers are the optimal level for each controllable factor. A) interaction effects for $25d$ observed samples, B) interaction effects for $37.5d$ observed samples, and C) interaction effects for $50d$ observed samples, }
\label{Interactioneffect}
\end{figure}

\subsection{Interaction Effects}
Next, we might be interested in investigating the interaction effects between the controllable and noise factors, as this could guide designers to make modeling decisions based on prior knowledge that they have of their experimental setup. In Figure~\ref{Interactioneffect} we have plotted the GAP averaged over all experiments that have the same controllable and noise factors for the different number of observed total samples (i.e., $d=\left\{ 25d, 37.5d, 50d \right\}$). 

What we can observe from Figure~\ref{Interactioneffect} is that there are few interaction effects between the controllable and noise factors. This can be concluded from the observation that the optimal levels for each controllable factor are the same. The exception to this are the following two interaction effects.
\begin{enumerate}
\item The total number of observed samples has an interaction effect with the initial number of replicates. Specifically, as the experimental budget increases, it becomes advantageous to increase the initial number of replicates.
\item The total number of observed samples has an interaction effect with the selected acquisition function. Specifically, for a small experimental budget, the PI is most appropriate, whereas for larger experimental budgets the UC bound becomes more appropriate. 
\end{enumerate}
The interaction effect between the total number of samples and the number of initial replicates suggests that replication is beneficial for  approximating the stochastic response surfaces. Specifically, when a larger experimental budget is available. Concerning the second interaction effect, we find that the acquisition function that places more emphasis on exploitation performs better with a larger experimental budget. This suggests, that most acquisition functions place too much emphasis on exploitation causing the sampling path to spend too many resources exploring the trivial region of the space of admissible input conditions. This is caused by the inflated uncertainty of the posterior predictive response as a consequence of the training data uncertainty.

\subsection{Recommendations}
From the previous results, we can make the following recommendations.
\begin{enumerate}
\item If your experimental budget is in the range of $15d$ to $40d$, then it is recommended to replicate the initial number of samples twice. In addition, for a larger experimental budget, more replicates might be required. Conversely, for an experimental budget of less than $15d$, it could be better to have no replicates. Although, this cannot be confirmed definitively from the performed study. 
\item For larger experimental budgets, it is recommended to start with a larger number of initial samples. Specifically, for less than $30d$ it is recommended to have an initial sample size of $5d$, whereas for larger experimental budgets $10d$ or more is recommended. 
\item For an experimental budget of less than $20d$ it is recommended to use the PI covariance function. For larger experimental budgets, it is best to use the UC interval with a relatively large value for $\pi$ (e.g., $\pi>5$).
\item The Mat\'{e}r covariance has a superior performance on all test problems compared to other covariance functions. 
\item If there is no explicit limitation on the experimental budget, then it is advised to use two initial replicates for all $10d$ initial sampling locations and use the Mat\'{e}r covariance in combination with either the PI or EI acquisition function. 
\end{enumerate}

\section{DISCUSSION}
In this section, we will briefly discuss some of the considerations that went into the performed study and the aforementioned observations.
\begin{enumerate}
\item In this study we have assumed a functional form for the experimental uncertainty, either the variance is homoscedastic, or it is proportional to the functional response. While in practice, more functional forms of the variance could exist, the set studied in the paper is significantly broad. For example, experimental uncertainty, a form of intrinsic uncertainty, is often constant or proportional to the response surface. 
\item We focused our study on normally distributed experimental uncertainty. Consequently, a designer should be careful of this when adopting the conclusion presented in this study when the experimental noise might be normally distributed. 
\item We did not study the acquisition of batches of samples \cite{Beek2021} and this could be important to the overall performance of the data acquisition process. While this is certainly a limitation of the presented study, it should be noted that our aim was to analyze low experimental setups with small experimental budgets. Consequently, sampling batches in this context is not practical. 
\item One assumption in the presented approach is that we have used the maximum likelihood approach to approximate the GP hyperparameters. As a consequence, the uncertainty in the hyperparameters is not considered when evaluating the posterior predictive distribution (this can be observed from the true function poorly fitting the 95\% confidence intervals in Figure~\ref{AQplots}). In addition, this also explains why most acquisition functions place more emphasis on exploitation. Consequently, it could be the case that using a Bayesian approach in the approximation of GP hyperparameters would be a more appropriate approach for the optimization of the stochastic response surface, especially with small experimental budgets. 
\end{enumerate}

\section{CONCLUSIONS}
In this paper, we have presented an experimental study in the modeling conditions associated with Bayesian optimization for noisy training data (e.g., physical experiments). In this study, we emphasized the scenario where only a small experimental budget is available, in which case stringent simplifying assumptions need to be made. Moreover, with this study, we aimed to identify what modeling conditions are most appropriate based on a priori knowledge on the nature of the problem (e.g., the magnitude of the experimental uncertainty, the functional form of the experimental uncertainty, and the form of the objective function). From this study, we have discovered that the Mat\'{e}r covariance function performed better than all studied alternatives (i.e., Gaussian, and power exponential). In addition, there is a strong correlation between the experimental budgets and many of the controllable modeling considerations. Specifically, the experimental budget has an interaction effect with the initial number of replicates, the initial number of samples, and the choice of acquisition function. Alternatively, the form of the objective function, the magnitude of the experimental uncertainty, and the form of the experimental uncertainty do have no interaction effects with the controllable modeling decisions. 

The work in this paper provides inspiration for future research directions. Specifically, it was found that conventional acquisition functions place too much emphasis on exploitation. This could be remedied by a Bayesian approach that accounts for the uncertainty in the hyperparameters of the emulator or through the development of problem-specific acquisition functions. More importantly, while many advanced methods for experimental design are available, physical experiments are often performed on input conditions that are selected based on the intuition of the designer. To make this process more systematic we explored the potential of adaptive sampling strategies, typically used in conjunction with simulation-based design, for physical experiments. However, as this involved simplifying assumptions, future research should be directed toward the discovery of new statistical methods for the optimization of noisy response surfaces in sparse data scenarios.

\section*{ACKNOWLEDGEMENTS}
The financial support from the School of Mechanical and Materials Engineering at University College Dublin is greatly appreciated. 

{\small
\bibliographystyle{unsrt}
\bibliography{main.bbl}

\begin{thebibliography}{10}

\bibitem{Forrester2009}
Alexander~I.J. Forrester and Andy~J. Keane.
\newblock Recent advances in surrogate-based optimization.
\newblock {\em Progress in Aerospace Sciences}, 45(1):50--79, 2009.

\bibitem{chaney2024}
Lindsay~E Chaney, Anton van Beek, Julia~R Downing, Jinrui Zhang, Hengrui Zhang,
  Janan Hui, E~Alexander Sorensen, Maryam Khalaj, Jennifer~B Dunn, Wei Chen,
  et~al.
\newblock Bayesian optimization of environmentally sustainable graphene inks
  produced by wet jet milling.
\newblock {\em Small}, 20(33):2309579, 2024.

\bibitem{hui2024}
Janan Hui, Haoyang You, Anton Van~Beek, Jinrui Zhang, Arash Elahi, Julia~R
  Downing, Lindsay~E Chaney, DoKyoung Lee, Elizabeth~A Ainsworth, Santanu
  Chaudhuri, et~al.
\newblock Biorenewable exfoliation of electronic-grade printable graphene using
  carboxylated cellulose nanocrystals.
\newblock {\em ACS Applied Materials \& Interfaces}, 16(42):57534--57543, 2024.

\bibitem{Beek2021}
Anton van Beek, Umar~Farooq Ghumman, Joydeep Munshi, Siyu Tao, TeYu Chien,
  Ganesh Balasubramanian, Matthew Plumlee, Daniel Apley, and Wei Chen.
\newblock Scalable adaptive batch sampling in simulation-based design with
  heteroscedastic noise.
\newblock {\em Journal of Mechanical Design}, 143(3), 2021.

\bibitem{Santner2003}
Thomas~J Santner, Brian~J Williams, William~I Notz, and Brain~J Williams.
\newblock {\em The design and analysis of computer experiments}, volume~1.
\newblock Springer, 2003.

\bibitem{Box1964}
George~EP Box and David~R Cox.
\newblock An analysis of transformations.
\newblock {\em Journal of the Royal Statistical Society: Series B
  (Methodological)}, 26(2):211--243, 1964.

\bibitem{Mukerjee2006}
Rahul Mukerjee and Chien-Fu Wu.
\newblock {\em A modern theory of factorial design}.
\newblock Springer, 2006.

\bibitem{Jin2005}
Ruichen Jin, Wei Chen, and Agus Sudjianto.
\newblock An efficient algorithm for constructing optimal design of computer
  experiments.
\newblock {\em Journal of Statistical Planning and Inference}, 134(1):268--287,
  2005.

\bibitem{Sobol1976}
I.M. Sobol.
\newblock Uniformly distributed sequences with an additional uniform property.
\newblock {\em USSR Computational Mathematics and Mathematical Physics},
  16(5):236--242, May 1976.

\bibitem{Igder2012}
AARA Igder, Ali~Akbar Rahmani, Ali Fazlavi, Mohammad~Hossein Ahmadi,
  Mohammad~Hossein Ahmadi~Azqhandi, and Mohammad~Hassan Omidi.
\newblock Box-behnken design of experiments investigation foradsorption of cd2+
  onto carboxymethyl chitosan magnetic nanoparticles.
\newblock {\em Journal of Mining and Environment}, 3(1):51--59, 2012.

\bibitem{Durakovic2017}
Benjamin Durakovic and Muris Torlak.
\newblock Experimental and numerical study of a pcm window model as a thermal
  energy storage unit.
\newblock {\em International Journal of Low-Carbon Technologies},
  12(3):272--280, 2017.

\bibitem{Lindquist1953}
Everet~Franklin Lindquist.
\newblock {\em Design and analysis of experiments in psychology and education.}
\newblock Houghton Mifflin, 1953.

\bibitem{Fisher1992}
Ronald~A Fisher.
\newblock The arrangement of field experiments.
\newblock In {\em Breakthroughs in statistics}, pages 82--91. Springer, 1992.

\bibitem{Binois2018}
Mickael Binois, Robert~B Gramacy, and Mike Ludkovski.
\newblock Practical heteroscedastic gaussian process modeling for large
  simulation experiments.
\newblock {\em Journal of Computational and Graphical Statistics},
  27(4):808--821, 2018.

\bibitem{Ankenman2008}
Bruce Ankenman, Barry~L Nelson, and Jeremy Staum.
\newblock Stochastic kriging for simulation metamodeling.
\newblock In {\em 2008 Winter Simulation Conference}, pages 362--370. IEEE,
  2008.

\bibitem{Plumlee2014}
Matthew Plumlee and Rui Tuo.
\newblock Building accurate emulators for stochastic simulations via quantile
  kriging.
\newblock {\em Technometrics}, 56(4):466--473, 2014.

\bibitem{Springenberg2016}
Jost~Tobias Springenberg, Aaron Klein, Stefan Falkner, and Frank Hutter.
\newblock Bayesian optimization with robust bayesian neural networks.
\newblock {\em Advances in neural information processing systems}, 29, 2016.

\bibitem{Williams2006}
Christopher~KI Williams and Carl~Edward Rasmussen.
\newblock {\em Gaussian processes for machine learning}, volume~2.
\newblock MIT press Cambridge, MA, 2006.

\bibitem{De2021}
George De~Ath, Richard~M Everson, and Jonathan~E Fieldsend.
\newblock How bayesian should bayesian optimisation be?
\newblock In {\em Proceedings of the Genetic and Evolutionary Computation
  Conference Companion}, pages 1860--1869, 2021.

\bibitem{Martin2005}
Jay~D Martin and Timothy~W Simpson.
\newblock Use of kriging models to approximate deterministic computer models.
\newblock {\em AIAA journal}, 43(4):853--863, 2005.

\bibitem{Shahriari2015}
Bobak Shahriari, Kevin Swersky, Ziyu Wang, Ryan~P Adams, and Nando De~Freitas.
\newblock Taking the human out of the loop: A review of bayesian optimization.
\newblock {\em Proceedings of the IEEE}, 104(1):148--175, 2015.

\bibitem{Jones2001}
Donald~R. Jones.
\newblock A taxonomy of global optimization methods based on response surfaces.
\newblock {\em International Conference on Machine Learning}, 21(4):345--383,
  2001.

\bibitem{Torn1989}
Aimo T{\"o}rn and Antanas Zilinskas.
\newblock {\em Global optimization}, volume 350.
\newblock Springer, 1989.

\bibitem{Huang2006}
Deng Huang, Theodore~T Allen, William~I Notz, and Ning Zeng.
\newblock Global optimization of stochastic black-box systems via sequential
  kriging meta-models.
\newblock {\em Journal of global optimization}, 34(3):441--466, 2006.

\bibitem{Frazier2008}
Peter~I Frazier, Warren~B Powell, and Savas Dayanik.
\newblock A knowledge-gradient policy for sequential information collection.
\newblock {\em SIAM Journal on Control and Optimization}, 47(5):2410--2439,
  2008.

\bibitem{Frazier2012}
Ilya~O Ryzhov, Warren~B Powell, and Peter~I Frazier.
\newblock The knowledge gradient algorithm for a general class of online
  learning problems.
\newblock {\em Operations Research}, 60(1):180--195, 2012.

\bibitem{Tao2021}
Siyu Tao, Anton Van~Beek, Daniel~W Apley, and Wei Chen.
\newblock Multi-model bayesian optimization for simulation-based design.
\newblock {\em Journal of Mechanical Design}, 143(11), 2021.

\bibitem{Hernandez2014}
Jos{\'e}~Miguel Hern{\'a}ndez-Lobato, Matthew~W Hoffman, and Zoubin Ghahramani.
\newblock Predictive entropy search for efficient global optimization of
  black-box functions.
\newblock {\em Advances in neural information processing systems}, 27, 2014.

\bibitem{Minka2001}
Thomas~Peter Minka.
\newblock {\em A family of algorithms for approximate Bayesian inference}.
\newblock PhD thesis, Massachusetts Institute of Technology, Cambridge, MA,
  USA, 2001.

\bibitem{Picheny2013}
Victor Picheny, Tobias Wagner, and David Ginsbourger.
\newblock A benchmark of kriging-based infill criteria for noisy optimization.
\newblock {\em Structural and multidisciplinary optimization}, 48:607--626,
  2013.

\end{thebibliography}
}

\end{document}